\documentclass{elsarticle}
\let\today\relax
\makeatletter
\def\ps@pprintTitle{%
    \let\@oddhead\@empty
    \let\@evenhead\@empty
    \def\@oddfoot{\footnotesize\itshape
         {} \hfill\today}%
    \let\@evenfoot\@oddfoot
    }
\makeatother
\usepackage[utf8]{inputenc}
\usepackage[T1]{fontenc}
\usepackage[english]{babel}
\usepackage{microtype} 
\usepackage{tabularx} 
\usepackage{booktabs} 

    \newcommand{\e}[1]{\mbox{\lstinline[basicstyle=\normalsize, language=OOSC2Eiffel]|#1|}}

    \usepackage{boogie}
    \usepackage{eiffel}
    \usepackage{listings}
    \usepackage{enumitem}
    \usepackage{hyperref}
    \usepackage{tabto}
    \usepackage{float}
    \usepackage{graphicx}
    \graphicspath{{./figures/}}
    \usepackage{longtable}
    \usepackage{tcolorbox}
    \usepackage[framemethod=TikZ]{mdframed}
    \usepackage{setspace}
    \usepackage{xurl}
    \mdfdefinestyle{MyFrame}{%
        linecolor=grey!50!white,
        outerlinewidth=2pt,
        roundcorner=20pt,
        innertopmargin=6pt,
        innerbottommargin=6pt,
        innerrightmargin=20pt,
        innerleftmargin=20pt,
        backgroundcolor=white}

    \newcommand{\mysec}[1]{section~\ref{#1}}
    \newcommand{\myfig}[1]{Fig.~\ref{#1}}
    \newcommand{\mytab}[1]{Table~\ref{tab:#1}}

    \setlist[itemize]{leftmargin=*, topsep=1pt, noitemsep}
    \setlist[enumerate]{leftmargin=*, topsep=1pt, noitemsep}
    \usepackage{caption}
    
\begin{document}
\begin{frontmatter}

\title{
Object-Oriented Requirements: a Unified Framework for Specifications, Scenarios and Tests
}

\author[inst2]{Maria Naumcheva}
\author[inst2]{Sophie Ebersold}
\author[inst4]{Alexandr Naumchev}
\author[inst2]{Jean-Michel Bruel}
\author[inst5]{Florian Galinier}
\author[inst3]{Bertrand Meyer}

\address[inst2]{IRIT, University of Toulouse, Toulouse, France}
  
\address[inst4]{Unaffiliated}          
\address[inst5]{Spilen Corporation, Toulouse, France}

\address[inst3]{Constructor University, Schaffhausen, Switzerland}

\begin{abstract}
A paradox of requirements specifications as dominantly practiced in the industry is that they often claim to be object-oriented (OO) but largely rely on procedural (non-OO) techniques. Use cases and user stories describe functional flows, not object types. 
To gain the benefits provided by object technology (such as extendibility, reusability, reliability), requirements should instead take advantage of the same data abstraction concepts -- classes, inheritance, information hiding -- as OO design and OO programs.

Many people find use cases and user stories appealing because of the simplicity and practicality of the concepts. Can we reconcile requirements with object-oriented principles and get the best of both worlds?

This article proposes a unified framework. It shows that the concept of class is general enough to describe not only “objects” in a narrow sense but also scenarios such as use cases and user stories and other important artifacts such as test cases and oracles.

Having a single framework opens the way to requirements that enjoy the benefits of both approaches: like use cases and user stories, they reflect the practical views of stakeholders; like object-oriented requirements, they lend themselves to evolution and reuse.
\end{abstract}

\end{frontmatter}

\section{Introduction} \label{introduction}

A good software system is an effective solution to a well-understood problem. As software engineering has progressed, it has become increasingly clear that achieving software quality involves achieving quality on both the solution side and the problem side: together with excellent design, implementation and project management techniques, a successful project requires an excellent description of the problem, known as the \textbf{requirements} of the system.

While a considerable body of knowledge exists about requirements engineering, the discipline as practiced in industry has not yet experienced the considerable progress that \textit{object-oriented} (OO) concepts, methods, languages and tools have brought to solution-side tasks. The purpose of this article is to help advance the state of the art in requirements engineering through the application of OO ideas, and to show that this approach subsumes other widely applied techniques such as use cases and user stories. 
The research questions we tackle in this paper are (i) how to specify OO requirements? (ii) how to unify them with scenarios?

The \textit{modeling power} of object technology has played a large part in its success for design and implementation, and can be even more useful for requirements. It comes in particular from the OO decision to define the architecture of systems on the basis of object types connected by  well-defined relations (``client'' and ``inheritance''), using structuring principles such as information hiding and Design by Contract. These notions are clear and simple, and help ensure that the structure of the software is closely related to the structure of the problem description. For example, a library-management system will have such classes as \e{LIBRARY}, \e{BOOK}, \e{PATRON} and \e{CATALOG},  directly reflecting concepts of the problem domain and their mutual relations. This property, known as \textit{direct mapping}, helps achieve goals of reliability, extendibility and reusability. It is even more useful for requirements, which are all about modeling external systems and their environments.

In practice, however, requirements engineering does not widely apply OO techniques. The phrase ``object-oriented requirements'' itself is used (sometimes as ``object-oriented \textit{analysis}''), but often to describe techniques such as use cases and user stories, which are procedural (the accepted term for ``non-object-oriented'').

The relationship between OO and procedural techniques can be one of complementarity rather than confrontation. The core observation is that data abstraction, the idea at the heart of OO technology, has so much power that some of it remains  untapped. One does not need to treat such techniques as scenarios, use cases, user stories and test scripts as independent of the OO framework or even antagonistic to it; they can instead find their natural place in it. The result is a unified approach to requirements engineering. which has the potential to bring to this discipline the same remarkable advances that have proved so beneficial to solution-side aspects of software development.
The first contribution of this paper is the evidence of scenarios limitations and how OO paradigm could address them; the following is the description of OO requirements.  The third contribution is a framework supporting OO requirements specification; the last one is the application of the approach to the case study of a Roborace, which can serve as a guide for practitioners who want to use the unified framework. 
These contributions allow requirements engineers to benefit from the advantages of conceptual consistency and unification of notations, and from the potential of scalability and extendibility of OO requirements.

Section \ref{commonApproaches} presents common approaches to requirements and their limitations.
Section \ref{OOfund} describes fundamental OO techniques, that form the basis of the Unified Approach.
Section \ref{OORQ} introduces the notion of object-oriented requirement.
Section \ref{Object-oriented requirements as the unifying framework} describes in details a unified framework, based on OO requirements.
Section \ref{application} constitutes a proof of concept by applying the framework to a real-case application.
Section \ref{related} compares the approach with related works.
Section \ref{discussion} describes motivation for applying OO techniques to requirements, limitations of the approach and of the case study, and future work directions.
Finally, section \ref{conclusion} summarizes the contributions and concludes the paper.

\section{Scenarios and tests}\label{commonApproaches}

Requirements in industry generally rely on different 
techniques, particularly use cases and user stories. In modern software development approaches, tests are viewed also as requirements artifacts. 

\subsection{Use cases} \label{use_cases}

Use cases have become one of the major formalisms for expressing requirements thanks to Ivar Jacobson’s work and his 1992 book \cite{ivarjacobson}. Their spread happened at about the same time as the spread of OO methods for programming and design. A use case describes a unit of requirements in the form of a possible scenario of user interaction with the system.

There are several use case notations; the examples in this article will rely on a notation due to Cockburn \cite{cockburn2000writing}.
\mytab{borrowUC} presents an example, related to a library system. It specifies a scenario whereby a library user borrows a book from the library, with such steps as placing the book on hold, then checking it out, and returning it by the specified deadline.
\newline

\begin{table}[htb!]
\centering
\small
\begin{tabular}{ |p{0.17\textwidth}|p{0.77\textwidth}| } 

 \hline
 Name & Borrow\_a\_book \\
  \hline
 Scope & System  \\ 
  \hline
 Level & Business summary  \\ 
 \hline
 Primary actor & Patron \\
 \hline
 Context of use & The patron wants to check out a book \\
 \hline
 Preconditions &  The book is available \\
 \hline
 Trigger & The patron finds in a library catalogue the book he wants to borrow and requests the system to place a hold on this book \\
 \hline
 Main success   & * The system changes the book status to on\_hold \\
scenario&
* The patron checks out the book\\
&
* The patron returns the book\\

\hline
Success guarantee & The patron has borrowed the book and returned it within the checkout duration.\\
\hline
Extensions
& A. The book is not available\\
 & *The system denies placing hold on the book \\
& B. The hold expires due to exceeding maximum hold duration. \\
& * The system changes the hold status to ``expired'' and the book becomes available\\
& C. The patron cancels the hold\\
& * The book status changes to available\\
& D. The patron does not return the book within the maximum check out duration\\
& * The book status changes to overdue\\
& * The patron returns the book\\

 \hline
Stakeholders 
& Patron (borrows a book) \\
and interests& Library personnel (enforces adherence to library policies)\\
 \hline
 
\end{tabular}
    \caption{Use Case description for \e{Borrow\_a\_book}}
    \label{tab:borrowUC}
\end{table}

The core part is the \textbf{main success scenario}, giving the sequence of steps of the use case. Additional possibilities of the notation, not used in the example, include:

\begin{itemize} \setlength{\itemsep}{0pt}
\item Steps consisting of several sub-steps, similar to calling routines in programming.
\item  Conditional steps, in which the use case follows either of two sub-scenarios depending on the outcome of a certain condition.
\end{itemize}

Such mechanisms suggest a strong analogy between a use case and an \textit{algorithm} or a program. The two concepts also have differences: a use case (and other kinds of scenario such as user stories, reviewed next) describes the interaction between a human actor and the system, whereas an algorithm or program is meant to be carried out by a computer.

The \textbf{level} entry characterizes the level of abstraction. A use case can describe a process at many levels, from the highest (a bird’s eye view of an overall business process, meant to be complemented by further use cases for the details) down to the detailed descriptions of the system’s actual operation.

A \textbf{precondition} is a limiting condition governing the applicability of the use case.

A \textbf{trigger} is an event that starts the use case.

Note the difference between the last two concepts: a precondition is a condition that must hold for the use case to be applicable, but does not by itself cause its execution; a trigger does. (The precondition is necessary, the trigger is sufficient.) 

The \textbf{success guarantee} characterizes the state resulting from successful execution of the use case, for the ``main success scenario''. In the terminology of mathematical software verification, from which the term ``precondition'' is borrowed, it would be called a ``postcondition''.  

An \textbf{extension} describes a departure from the ``main success scenario''.
Extensions serve two separate purposes:
\begin{itemize} \setlength{\itemsep}{0pt}
\item  As in the example, an extension can specify an alternative to the “main success scenario”, to be applied when that scenario hits a condition that prevents it from proceeding normally.
\item Extensions also support reuse of elements common to several use cases, which can then be divided into a base use case, covering the common elements, and specific extensions.
\end{itemize}
The use case lists, towards the beginning, the \textbf{main actor} responsible for carrying out instances of the scenario. 
It concludes with a list of \textbf{stakeholders}: others who may be affected.

\subsection{User stories} \label{user_stories}

A use case is a complete path taken by an actor through the system. User stories  \cite{usecase20} also express a typical interaction with the system, but at a much smaller level of granularity. They play an important role for requirements in \textbf{agile methods}, which promote incremental program construction: the basic agile development iteration involves a developer picking the next item from a list of functions to be implemented, implementing it, and moving on to the next one. A use case is generally too complex for such atomic units of development.

The standard format for a user story includes three elements:
\begin{itemize} \setlength{\itemsep}{0pt}
\item A role (``\textbf{As a...}''), corresponding to the main actor of a use case.
\item A desired function (``\textbf{I want to...}''), part of the system’s behavior.
\item A business purpose (``\textbf{so that...}''), corresponding to one of the goals of a system. 
\end{itemize}

An example user story in a library system is: 

\smallskip

{\textbf{As a}\textit{ patron,} \textbf{I want to} \textit{check out a book }\textbf{so that}\textit{ I can read it at home}}.

\vspace{\baselineskip}

An alternative way of expressing it is in tabular form as presented in \mytab{US}.
\newline

\begin{table}[htb]
    \centering
    \small
\begin{tabular}{ |p{0.25\textwidth}|p{0.5\textwidth}| } 
 \hline
 Role &  Patron \\
  \hline
 Desired function & Check out a book \\ 
  \hline
 Business purpose & Read at home \\ 
 \hline
 \end{tabular}
    \caption{User Story Example}
    \label{tab:US}
\end{table}

\subsection{Use case 2.0 and use case slices} 
\label{usecase2} 
Use Case 2.0 \cite{use_case_slice_enacted}, a revision of the original use case methodology, combines ideas of use cases and user stories through the notion of use case slice.  
A use case slice is a selected part of a use case, which is also a unit of testing (usable in a test-driven design methodology). Breaking down a use case into slices makes it possible to consider it at different phases of development and to build the system incrementally.

As an example, a slice of the use case \textit{Borrow\_a\_book} is \textit{Overdue\_checkout} (\mytab{Overdue}). 
It outlines its narratives as a set of flows: the basic flow of \textit{Borrow\_a\_book} and one of its extensions (Alternative Flow D), ``\textit{The patron does not return the book within the maximum check out duration.}''
\newline
\begin{table}[htb]
    \centering
    \small
\begin{tabular}{ |p{0.18\textwidth}|p{0.75\textwidth}| } 
 \hline
 Name &  Overdue\_checkout \\
  \hline
 State & Scoped  \\ 
  \hline
 Priority & Should  \\ 
 \hline
Flows & BF: ``\textit{Borrow a book}''  + AF D: ``\textit{The patron does not return the book within the maximum check-out duration}''\\
 \hline
Tests & The checkout duration exceeds the limit\\
 \hline
 Estimate & 3/20 \\
 \hline
\end{tabular}
    \caption{Use case slice description of \textit{Overdue\_checkout}}
    \label{tab:Overdue}
\end{table}

The standard format of a use case slice includes: 
\begin{itemize}
    \item A \textbf{name}, used to track it through the development cycle.
   \item A \textbf{state}, with possible values \textit{scoped}, \textit{prepared}, \textit{analysed}, \textit{implemented} or \textit{verified}.
   \item A \textbf{priority}, expressed by the MoSCoW acronym: Must, Should, Could, Would. 
   \item References:  \textbf{flows}
   and  \textbf{tests}
   \item An \textbf{estimate} of the work needed to implement the slice.
\end{itemize} 

\subsection{Unit tests} \label{unit_tests}

An important development of software engineering in the past two decades, fostered in particular by agile methods, has been the recognition of test cases as essential artifacts of the software process. The various ``xUnit'' frameworks, where ``x'' represents a target programming language (for example, JUnit for Java), describe a test case in the form of a small program element listing a condition to be tested and an ``oracle'' specifying the correct expected result.

In a traditional, ``waterfall'' view of software engineering, requirements and test cases play entirely different roles and appear at opposite ends of the lifecycle: requirements at the beginning, tests at the end (in a final  ``verification and validation'' step, often called V\&V). The only relationship between them is that a test case, devised in a late stage of the project, will typically exercise a particular requirement, specified much earlier.

With modern views of software development, iterative and incremental, the concepts move closer to each other. In the extreme view (coming from the ``Extreme Programming'' approach to development) of \textit{test-driven development} \cite{beck2003test}, unit tests actually become the requirements, pushing aside more traditional forms of requirements. Even in less radical approaches, tests share many of the properties expressed as use cases or user stories.

The following example is a Java test for the example library system (it corresponds to a requirement given in \mysec{specification_drivers} below). 

\begin{lstlisting}[language=Java, showstringspaces=false]
public class HoldingAvailableBooksTest {
  private Book b; 
  private Patron p1, p2; 
  private LibraryBranch lb;
  private Library l;
  
  $@$BeforeEach
  public void setUp() {
    b = new Book("Crime and Punishment", "Fedor Dostoyevsky", 
                "978-1703766172");
    p1 = new Patron("Ted"); p2 = new Patron("Fred");
    lb = new LibraryBranch("Squirrel Hill");
    l = new Library("Carnegie Library of Pittsburgh");
    l.addBranch(lb); l.addPatron(p1); l.addPatron(p2);
  }
  
  $@$Test
  public void testHolding() {
    l.placeBookOnHold(b, p1, lb); 
    l.placeBookOnHold(b, p2, lb);
    assertTrue(l.bookIsOnHold(b, p1, lb)); 
    assertFalse(l.bookIsOnHold(b, p2, lb);
} }
\end{lstlisting}

The \e{setUp()} method instantiates concrete input objects for the test. Then \e{testHolding()} executes the method under test and checks correctness of the objects in the resulting state.
The \e{$@$BeforeEach} annotation directs the unit testing framework to execute the \e{setUp()} method before running any tests, decorated with \e{$@$Test} annotations.

The role of tests extends beyond requirements (the focus of this article); note in particular the widely accepted role of the \textit{regression test suite}, which collects all tests run on versions of a system since the project’s inception. Another idea that has gained wide acceptance under the influence of agile methods is a milder form of Test-Driven Development: the rule that every addition to the code must also include a new test (even if we avoid misconstruing that test for a specification). Some agile approaches go so far as to accept only code and tests are legitimate artifacts.

\subsection{Benefits and limitations of scenarios and tests for requirements}
\label{Limitations}

The techniques reviewed above have gained front-row seats in modern software development. Use cases and user stories are important as checks on requirements, to verify that the requirements, expressed in any form, do cover the most important cases. They are not, however, a substitute for these requirements.

A good mathematical textbook presents, along with every important theorem, examples of its consequences. A good physics textbook presents, along with every important property, examples of its applications. They help make the theorem or property more concretely understandable. No one, however, would accept the examples as a replacement for the theorem. Use cases are to a requirement what an example is to a theorem and a test to a specification.

A scenario or test describes only one path of system usage; the number of actual paths is, for any significant system, staggeringly large or theoretically infinite. Requirements should specify the system’s behavior in all cases, not just the inevitably small subset of cases that have been foreseen.

The difficulty of software and particularly of software requirements comes from the myriad of variants that a system must be prepared to handle. One of the prime goals of requirements is to help ensure the system’s correctness. In practice, the most serious bugs (violations of correctness) do not generally arise from the common example scenarios typically included in use cases, user stories and unit tests, since the testing phase will naturally focus on them; they arise out of less obvious cases, which such examples often leave out. Only by trying for a more abstract form of specification, covering all cases, can one expect to identify such problematic cases.

Even use cases, the most advanced form of scenario, suffer from this limitation. A use case describes only one path of interaction with the system, or a set of related cases if it takes advantage of conditional branching and extensions. In practice, use cases tend to focus on the most desirable cases, in particular the ``main success scenario'' and possibly some of its variants as illustrated earlier in section \ref{use_cases}) . While one can add extensions covering such situations, or write new use cases to address them, there are so many combinations for any non-trivial system that it is impossible to include all such extensions, or even a representative subset. These observations also apply to user stories and tests.

Scenario-based techniques suffer from another limitation: their exaggerated reliance on time ordering, which often leads to overspecification. A use case will often specify a sequential contract (step B must occur after step A) when in fact a logical constraint would suffice (step B requires a certain condition, which step A ensures). Turning logical constraints into sequential ones is often too restrictive. Section \ref{logical} discusses this point further. (Also note Glinz's critical analysis of use cases \cite{Glinz}.)

The preceding analysis leaves two questions open:
\begin{itemize}
    \item Is a more general form of requirement available, focusing on abstract properties rather than examples?
    \item Can we express use cases, user stories and tests in the same framework as those requirements?
\end{itemize}

For answers, we may turn to the object-oriented mode of specification.

\section{Object-oriented fundamentals} \label{OOfund}

To reach the level of abstraction that scenarios do not provide, we look into object-oriented techniques, which have been widely applied to programming and design.

\subsection{OO principles}
The core OO concepts are listed below. 
\begin{itemize}
\item \textit{Type-based modular decomposition}: capture the properties of the key ``things'' manipulated by a system, better called \textbf{objects}, by abstracting them into types, or \textbf{classes}, and define the modular structure of a system as a set of interconnected classes.
\item \textit{Data abstraction}: describe each class not by implementation properties (what the corresponding objects ``are'') but by the applicable operations, or \textbf{features} (what the objects ``have'').
\item \textit{Contracts}: in such descriptions, include not only structural properties, such as the types of arguments handled by each operation, but also abstract semantic properties (preconditions, postconditions, class invariants).
\item \textit{Inheritance}: organize classes into taxonomies to take advantage of common traits. Techniques of polymorphism and dynamic binding, which follow from inheritance, enhance the architectural quality of OO systems.
\end{itemize}

The main benefits pursued by these techniques are extendibility (ease of modifying software, by keeping the various elements of a system isolated from changes in others), reusability (ease of applying elements of a system to the development of a new system) and reliability (correctness of solutions, robustness in the presence of errors, and security). These goals are just as desirable for requirements as they are for design and programming.

The next subsections examine some of the principal OO concepts in more depth, with an eye on their application to requirements. The examples use the Eiffel notation, which includes built-in mechanisms of Design by Contract and was designed to cover requirements and design in addition to programming.

\subsection{Classes and their mutual relations}
The central concept of object-oriented approaches to system modeling and structuring is the class, defined as a system unit specifying a type of objects with the associated operations and their properties. A class has operations of three kinds, which we may illustrate through the example of a class describing the concept of book in a library system:
\begin{itemize} 
\item Queries, providing information about the book: is\_available, isbn, author.
 \item Commands, to update the corresponding objects: hold, checkout.
 \item Creators (or ``constructors''), yielding objects of the type from other information, for example a book object defined by an author and title.
\end{itemize}

OO system descriptions are simple, relying on classes and only two relations between these classes: client and inheritance.
For example, operations on a book (see the use case presented in \mysec{use_cases}) involve, among other objects, a patron and a  library branch. The class representing a book will be a \textbf{client} of the classes representing the other three concepts.

\begin{figure}[!htbp]
\centering
\includegraphics[scale=0.5]{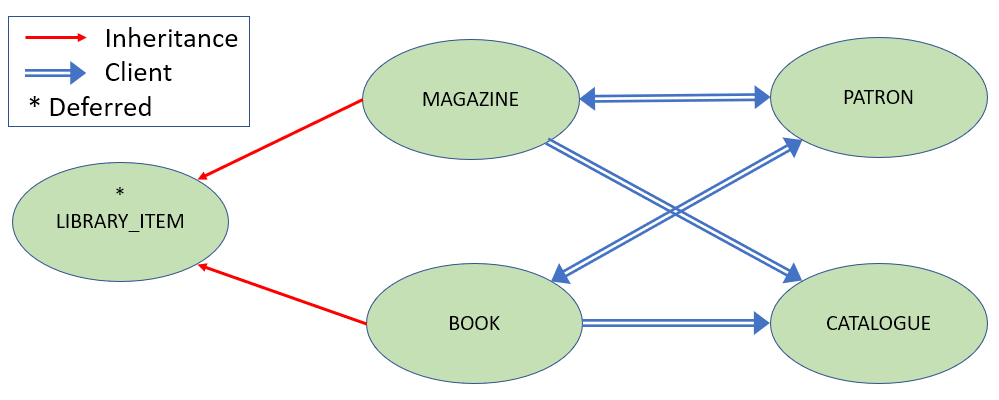}
\caption{Examples of the two kinds of inter-class relations}
\label{Inheritance_graph}
\end{figure}

A class \textbf{inherits} from another if it represents a specialized or extended version of the other’s concept. 
For example book and magazine both belong to the general category of library item, which can be represented by a class \e{LIBRARY\_ITEM}.
\e{BOOK} and \e{MAGAZINE} are ``descendants'' of
\e{LIBRARY\_ITEM} through the inheritance relation shown in  \myfig{Inheritance_graph}. 

\subsection{Deferred classes}

In an OO program, most of the commands and queries of a class are ``effective'', meaning that the class specifies an algorithm for their implementation. 
A class whose features are all effective is itself called an \textbf{effective} class. 
In some cases, however, including classes describing requirements, it is useful to define features without providing an implementation. 
Features specified but not implemented are called \textbf{deferred}; and a non-effective class (one that has at least one deferred feature) is a deferred class.
In requirements for the library example, \e{LIBRARY\_ITEM} will most likely be a deferred class (as marked by ``*'' in \myfig{Inheritance_graph}) since it describes an abstract concept with several possible concrete realizations. 

In requirements, declaring classes and features deferred makes it possible to specify that some functionality must exist while staying away from any consideration of implementation and hence from the risk of \textit{overspecification}. 

Even without an implementation, it will be possible in such cases to specify abstract properties of the desired behavior thanks to contracts, as discussed in \mysec{contracts} below.
It might seem that in requirements all classes and features should be deferred, but that would be too restrictive. 
As we will see in \mysec{Object-oriented requirements as the unifying framework}, requirements classes can specify scenarios, such as use cases or tests; such classes and their features will, in general, be effective since they prescribe precise sequences of operations.

\subsection{Contracts} 
\label{contracts}
Information hiding implies that modules of a system’s description (in OO modeling, classes) can refer to others in terms of their abstract specification. To be useful, such specifications should not just be structural, giving the types of arguments and results of the operations in a class, but also semantic, describing the abstract properties of these operations and the class as a whole. Design by Contract techniques \cite{OOSC} provide this semantic specification for both operations (preconditions and postconditions) and classes (class invariants).

As an example, the class extract below expresses the requirement that \textit{a public library allows patrons to place books on hold at its various library branches}:

\begin{lstlisting}[language=OOSC2Eiffel]
class LIBRARY feature
  place_book_on_hold (b: BOOK; p: PATRON; lb: LIBRARY_BRANCH)
    require
        has_patron (p)
        has_branch (lb)
    do -- Future implementation
    ensure
        book_is_on_hold (b, p, lb)
    end
end
\end{lstlisting}

The \textbf{require} clause introduces a precondition, and the \textbf{ensure} clause a postcondition.

Contracts have many applications, including as documentation of the software and as a guide to both exception handling and the proper use of inheritance. When applied to code, they also provide a systematic approach to testing and debugging (in a development environment that provides the ability to evaluate contract elements at run time). The application of most direct relevance to requirements, as illustrated by the \e{LIBRARY} example sketch above, is to model the System or Environment by providing not only structural properties (object types and the applicable operations) but also through precise semantics (the properties of these types and operations). 

Such semantic models lend themselves to automatic verification. Beyond run-time monitoring of contracts, which only applies to executable code, classes equipped with contracts can be verified using automatic tools such as JML \cite{leavens1999jml} and the AutoProof verifier for Eiffel \cite{tschannen2015autoproof}.

\subsection{Specification drivers}
\label{specification_drivers}
The object-oriented style of modeling has modular units (classes) each organized around one type of objects. Correspondingly, contract techniques apply within a class, making it possible for example to express the requirement that ``\textit{after a patron returns a book, it is considered available}'' as a postcondition of an operation \e{return} in a class \e{BOOK}.

Some properties apply to several objects, possibly of different types. An example is ``\textit{available books can be placed on hold by only one patron at any given time}''. 

While the OO style would seem to break down in such cases, it actually handles it in a simple way through the introduction of “specification drivers” \cite{naumchev2016complete}. The idea (generalizing techniques already present in some classic Design Patterns such as Visitor \cite{design_patterns}) is to express such cross-object properties through classes designed specifically for specification purposes.

In the last example, we may use the following specification-driver assertion, which describes a generic scenario of using the relevant features and specifies its effect through the postcondition:

\begin{lstlisting}[ language=OOSC2Eiffel]
holding_available_books (b: BOOK; p1, p2: PATRON; lb: LIBRARY_BRANCH; l: LIBRARY)
    require
        b.is_available; p1 /= p2
        l.has_patron (p1); l.has_patron (p2); l.has_branch (lb)
    do
        l.place_book_on_hold (b, p1, lb)
    l.place_book_on_hold (b, p2, lb)
    ensure
        l.book_is_on_hold (b, p1, lb)
        not l.book_is_on_hold (b, p2, lb)
    end
\end{lstlisting}

Specification drivers retain the OO style of requirements specification but make it more general by covering arbitrary properties, not necessarily expressible within a single class of the original OO model. 

\section{Object-oriented requirements} 
\label{OORQ}
The scope of object-oriented concepts as described in the previous section is broad enough to encompass all tasks of software construction, from requirements to design, implementation of course, and even verification. Of direct interest in this article is the application to requirements.

\subsection{OO requirements basics}
Discussing a system as a set of object types (classes) characterized by the applicable operations yields a number of potential benefits, which have been widely recognized in the programming world but can apply to requirements as well. They include the following:
\begin{itemize} \setlength{\itemsep}{0pt}
\item Stability through the evolution of the software. 
Changes affect individual modules, not the architecture.
\item Information hiding. With classes, we may declare, as part of the specification of a class, that some of the properties are for internal use only, within the class, and not accessible outside; limiting the effect of changes.
\item Reuse. If we are trying to apply the results of one project to another, reusing individual operations will generally not work. Operations such as “place a hold on a book”, “checkout a book”, “cancel a checkout” and others are closely connected. The notion of a book (as a class which includes all these operations) is a more realistic unit for reuse.
\item Classification. Inheritance makes it possible to describe new classes as extensions or specializations of existing ones, without repeating common properties. Along with information hiding, inheritance is a key tool in harnessing system complexity.
\item Modeling power (``direct mapping principle'', as noted in section \ref{introduction}).
OO concepts can yield system descriptions that are clear and intuitive, since the notions of object, class and inheritance are easy to grasp. Some classes have immediately tangible counterparts in the system environment. They are called “Environment classes”, Concrete and Abstract, in the classification presented in a subsequent section (\ref{modeling_enironment}).
\item Abstraction. Deferred classes and features make it possible to specify the presence of certain types of objects and of certain operations on them, without giving implementation details (but with abstract behavioral properties through contracts).
\end{itemize}

\subsection{Modeling the system and its environment}
\label{modeling_enironment}
Requirements in their general sense involve four aspects (called the “Four PEGS” in \cite{meyer_requirements_handbook}): Project (human effort to produce a system), Environment (the material or virtual reality in which the system will operate), Goals (objectives set by the organization), and System (the set of functionality elements that will be provided). OO modeling is applicable to all four aspects.

The application to the System part is the traditional focus of OO ideas. When the project reaches the design and implementation stages, System classes will include both concrete classes describing implementation elements and more abstract design classes, corresponding for example to design patterns such as Visitor or Observer.

At the requirements stage, it is not too early to use classes to identify the major components of the system and describing them through classes. Many of these classes will be deferred (abstract), but they may already include operational elements, including specification drivers and (as explained below) scenarios.

Of particular relevance are Environment classes; OO modeling is indeed appropriate to describe components of the environment. A typical example is the description of interfaces to and from other systems, through interface classes describing such elements as sensors, actuators, buttons and control panels. The corresponding objects are not part of software under development, but they are directly monitored or controlled by the software.

Contract elements are particularly useful here, to describe delicate properties of the environment. Such properties include:

\begin{itemize}
\item Constraints: conditions imposed by the environment, such as a physical limit (in a cyber-physical system) or a legal obligation (for a business system).
\item Effects: changes to the environment produced by the system (such as the triggering of an actuator, or a change in payroll processes). 
\item Assumptions (posited properties of the environment, making the system’s construction easier).
\item Invariants (properties that are both constraints and effects, as they can be assumed but must be preserved).
\end{itemize}

\subsection{An example OO specification}
\label{class_book}
The class text below is an example of how we can use an object-oriented formalism to describe requirements specifications, independently of any design or implementation consideration. It describes a notion of book in a library system. The following section will discuss some of its most important features distinguishing it from other types of specifications such as use cases.

\begin{lstlisting}[ language=OOSC2Eiffel]
class BOOK feature
      -- Boolean queries (is_available initialized to True;
      --                  is_on_hold, is_checked_out initialized to False)
    is_available, is_on_hold, is_checked_out: BOOLEAN 
    place_hold (patron: PATRON)
	    -- Place a hold on a book. Set is_on_hold 
        require
            is_available 
        deferred 
        ensure
            is_on_hold 
            not is_available 
        end
    checkout (patron: PATRON)
        -- Check out the book 
        require
            is_on_hold 
        deferred 
        ensure
            is_checked_out 
        end
    return 
        -- Return the book to the library 
        require
            is_checked_out 
        deferred 
        ensure
            is_available
	end 
invariant
    is_on_hold implies not is_available 
    is_checked_out implies not is_available 
    is_checked_out implies not is_on_hold 
    is_available implies not is_checked_out
end

\end{lstlisting}

\section{Object-oriented requirements as the unifying framework}
\label{Object-oriented requirements as the unifying framework}
The object-oriented approach is a structuring discipline, which models systems — at all levels: requirements, design, implementation — as collections of classes equipped with contracts and related to each other through client and inheritance links. This framework is general enough to encompass all aspects of requirements and provides room for the various non-OO techniques (scenarios and tests) reviewed earlier. The following sections show how to use an OO model as such an all-encompassing host for various applications. 

\subsection{Logical rather than sequential constraints} \label{logical}

A distinctive feature of the above \e{BOOK} class sketch is its reliance on logical constraints (through contracts) in lieu of a strict specification (or overspecification) of sequencing constraints.

OO techniques avoid premature \textit{time-ordering decisions}. While it is possible for an OO specification to express a time-ordered scenario such as a use case, object technology also supports a more general and abstract specification style, based on contracts.

The comparison of class \e{BOOK} with the use case specification of books in section \ref{use_cases} provides a good illustration. The use case version specifies the order in which operations will get executed; for example, in the ``main success scenario'':

\begin{itemize}
\item The system changes the book status to on\_hold
\item The patron checks out the book
\item The patron returns the book
\end{itemize}

Enforcing such an ordering specification at the level of requirements is often a premature decision. In reality, the order of the steps is not cast in stone. Using a preset ordering is convenient to describe \textit{desirable} scenarios, or more generally the \textit{expected} ones. But what happens in life is not always what we hope for, or expect. What if the customer returns a damaged book? Should the book not remain unavailable until it is repaired?

To specify scenarios that depart from the standard ones, we saw that it is possible to use \textbf{extensions}. But this solution does not scale. Writing ever more use case extensions to cover all such situations leads to an explosion of special cases which soon becomes be intractable. In practice, it is possible to write use cases to cover the most common scenarios, but they are only a small subset of the possible ones, in the same way that, in programming, tests can only cover a minute subset of possible inputs.

To get out of this predicament, we note that while constraints between the operations do exist, it is often more general and effective, instead of \textit{timing} constraints, to rely on \textit{logical} constraints. 

The preceding example scenario provides a good illustration. As a specification, it is trying to express a few useful things; for example, the patron should place a hold on a book before checking it out. But it states them in the form of a strict sequence of operations which does not cover the wide range of possible scenarios.

In the same way, a user story such as ``\textit{As a patron, I want to check out a book so that I can read it at home}" describes the interaction between a human actor and the system,. Describing a few such scenarios is useful as part of requirements elicitation, but to express the resulting requirements it is more effective to express the logical constraints. 

Class \e{BOOK} (from section \ref{class_book}) specifies these logical constraints in the form of contracts. Notice the interplay between the preconditions and postconditions and the various boolean-valued queries they involve (\e{is_available}, \e{is_on_hold}, \e{is_checked_out}). It is possible to specify a strict order of operations $o_1$, $o_2$, $...$, as in a use case, by having a sequence of assertions $p_i$ such that operation $o_i$ has the contract clauses \textbf{require} $p_i$ and \textbf{ensure} $p_i+1$; but assertions also make it possible to specify a much broader range of allowable orderings. Logical constraints are more general than sequential orderings.

The specific sequence of actions described in the use case (“Main scenario”) is compatible with the logical constraints: one can check that in the sequence

\begin{lstlisting}[language=OOSC2Eiffel]
        -- The following is the "Main scenario":
place_hold (patron: PATRON)
checkout (patron: PATRON)
return (patron: PATRON)

\end{lstlisting}

the postcondition of each step implies the precondition of the next one (the first has no precondition). Prescribing this order strictly is, however, overspecifying. For example it may be possible to perform additional operations between \e{place_hold} and \e{check_out}.

The contract-based specification does cover the fundamental sequencing constraints; for example, the pre- and postcondition combinations imply that investigation must come before evaluation and that resolution must be preceded by either negotiation or imposition. But they avoid the non-essential constraints which, in the use case, were only an artifact of the sequential style of specification, not a real feature of the problem \cite{Petre2013}.

\subsection{Integrating scenarios into an OO model}
\label{scenario_classes}
The preceding discussion shows how a specification through classes and their contracts beats a scenario-style specification in precision and generality. Use cases, user stories and other scenarios do retain their attractive features mentioned in section \ref{commonApproaches}, in particular their intuitive appeal to stakeholders, who can relate them easily to business processes, and their value as ways to validate the completeness of specifications.
There is no need, however, to sacrifice scenarios -- in particular use cases, retained as example for this discussion -- in an OO approach. A use case is simply a certain pattern of exercising the features (operations) of one or more classes; it can easily be expressed as a \textit{routine} (method) of an appropriate class.

Here for example is the expression of the use case from \ref{use_cases} as a routine that calls features of the class \e{BOOK}:

\begin{lstlisting}[language=OOSC2Eiffel]
class BOOK feature
...
    borrow_and_return_book (p: PATRON, lb: LIBRARY_BRANCH)
        require
            book_is_available: is_available
        do
            place_hold (p)
            checkout (p)
            return (p)
        end
end

\end{lstlisting}

This use case is simply a routine, calling features of the appropriate class.

Where should such use case routines appear? Two possibilities are available:
\begin{itemize}
\item In the case of a use case characterizing a single data abstraction and applying to a single object, such as one book, it can be expressed as a routine of the corresponding class, in this case \e{BOOK}. Then it simply describes a specific behavior of the instances of the class, expressed in terms of the more fundamental operations of the class.
\item A more general solution, and the one that fits the case of a use case involving several objects of possibly different types (hence, routines in different classes), is to group use cases into a separate class.
\end{itemize}

In both cases, a use case is an addition to one or more data abstractions from the rest of the requirements, intended to illustrate specific ways of using its features. The notion of  ``specification driver'' (section \ref{specification_drivers}) covers such specification elements exercising the features of one or more model classes. It was originally introduced for verification purposes (proofs and tests); we may view use-case class as an application of the concept to requirements specification. 

In accordance with OO principles, a use-case class should include not just one use case but a group of logically related use cases, exercising features of other requirements classes. (The idea of describing a single use case as a class is not new; see for example \cite{UML}. The use case classes described here are at a higher level of abstraction, covering a whole set of related behaviors, all pertaining to one or more data abstractions covered by other requirements classes.)   

Here is an example of such a use-case class (a collection of specification drivers) exercising features of classes \e{BOOK} and \e{LIBRARY}: 

\begin{lstlisting}[language=OOSC2Eiffel]
class LIBRARY_BOOK_USAGE feature
    borrow_and_return_book  
       do  ... as given above ... end
    decommission_book
        do ... Specification of decommission use case ... end
    renew_book
        do ... end
    -- Other use cases
end

\end{lstlisting}

To avoid any confusion, note that the idea is not to describe \textit{one} use case as a class or the corresponding object; that possibility is well-known, and not surprising since any well-defined abstraction can be modeled as an object.
A use class such as \e{LIBRARY_BOOK_USAGE} gathers a \textit{set} of important behaviors connected with an abstraction (or a group of abstractions) from the OO model of the environment and system. Such a set of related behaviors is in its own right a relevant abstraction in the OO model.

\subsection{Relation of OO requirements to test cases}

Test cases fit in the general OO framework just as use cases do. In modern ``xUnit" approaches to software testing (\mysec{unit_tests}), the basic scheme is already there since, as noted in the earlier discussion, these frameworks require writing test cases as routines. The earlier example, \e{HoldingAvailableBooksTest}, takes the form of the class \e{HOLDING_AVAILABLE_BOOKS_TEST} depicted below. 
\begin{lstlisting}[language=OOSC2Eiffel]
class HOLDING_AVAILABLE_BOOKS_TEST feature
  test_holding
    local
        b: BOOK
        p1, p2: PATRON
        lb: LIBRARY_BRANCH
        l: LIBRARY
    do
      create b.make("Crime and Punishment", "Fyodor Dostoyevsky", "978-1703766172");
      create p1.make("Ted"); create p2.make("Fred");
      create lb.make("Squirrel Hill");
      create l.make("Carnegie Library of Pittsburgh");
      holding_available_books (b, p1, p2, lb, l)
    end
    ...
end
\end{lstlisting}

As suggested by the final `...', the given routine \e{test_holding} does not need to be the only in its class. Having one testing routine per testing class is, in fact, the typical xUnit style as used in practice, but this practice misses the advantages of OO modularization. (It is a general rule of OO methodology that a class with just one routine is a ``design smell'', a sign of probably bad OO design.) In the context defined by the present discussion, the testing harness for a system should consist of a set of testing class, each exercising some features of a system or environment class (or of a closely related group of such classes).

\begin{sloppypar}
Like use cases, such test classes are example of ``specification drivers''. \e{HOLDING_AVAILABLE_BOOKS_TEST}, containing one or more routines related to book borrowing, is a typical example. 
\end{sloppypar}

\subsection{Premature design?}

An object-oriented requirements specification, as presented, uses an object-oriented notation, borrowed from an OO programming language. This notational resemblance may give the wrong impression of a specification involving premature choices of design and implementation. Such criticism is not justified; in fact, a proper OO requirements specification is more abstract, and less prone to overspecification, than a use-case or other scenario-based form of requirements.

The classes used in an OO specification are purely descriptive; they specify concepts (in particular, system and environment concepts) in an abstract way, using the OO style of specifying object types through the applicable operations and their abstract properties (contracts). In contrast, use cases are of a sequential, operational nature, presenting a risk of premature design, particularly as they make it tempting to write programs following the same ordering patterns, justified or not. 

While free from design and implementation considerations, the classes written for requirements purposes are still classes and can be expressed in an object-oriented language that also supports design and implementation classes. The benefit here is to avoid harmful changes of concepts and notations when going through successive steps of the software lifecycle. This approach is known as seamless development. One of its consequences is reversibility: having everything expressed in the same notation makes it easier to update the requirements at any stage in the project, even deep into design, implementation or verification.

\subsection{Applying the framework}

While the systematic description of a comprehensive approach to requirements specification falls beyond the scope of this article, we may build on the \e{BOOK} example to obtain an outline of the general process of object-oriented requirements specification:

\begin{enumerate}
    \item 
    Express the fundamental abstractions in the form of requirements classes.
    \item
    Express the fundamental constraints in the form of logical properties: invariants for these classes as well as preconditions and postconditions for their features (operations).
    \item
    Express typical usage scenarios through use cases or user stories. (Unlike the previous two, this task does not make any attempt at exhaustiveness, since examples can only cover a minute fragment of all possibilities; instead, it concentrates on the scenarios of most interest to stakeholders, and those most likely to cause potential issues or bugs.)
    \item
    As a consistency check, ascertain that the scenarios (item 3) preserve the logical properties (item 2). Update the logical properties if needed.
    
\end{enumerate}

\noindent More generally, the combination of an object-oriented approach to structure the requirements (1), equipped with invariants (2) as well as other forms of contracts (preconditions, postconditions), with use cases to illustrate the requirements through examples of direct interest to stakeholders (3) and shown to preserve the invariants (4) provides, as sketched here for the Roborace system, a promising method for obtaining correct and practically useful requirements.

\section{A case study -- the Roborace software} \label{application}

This section illustrates the application of the suggested approach to a real-world case study: the Roborace \cite{roborace_site}. 
The code excerpts used in this section are available at \cite{roborace_spec}. 
Although the complete requirements specification is beyond the scope of the article, we illustrate the implementation of the key concepts of the framework.

\subsection{Roborace: informal description}

\begin{figure}[!htbp]
\centering
\includegraphics[scale=0.1]{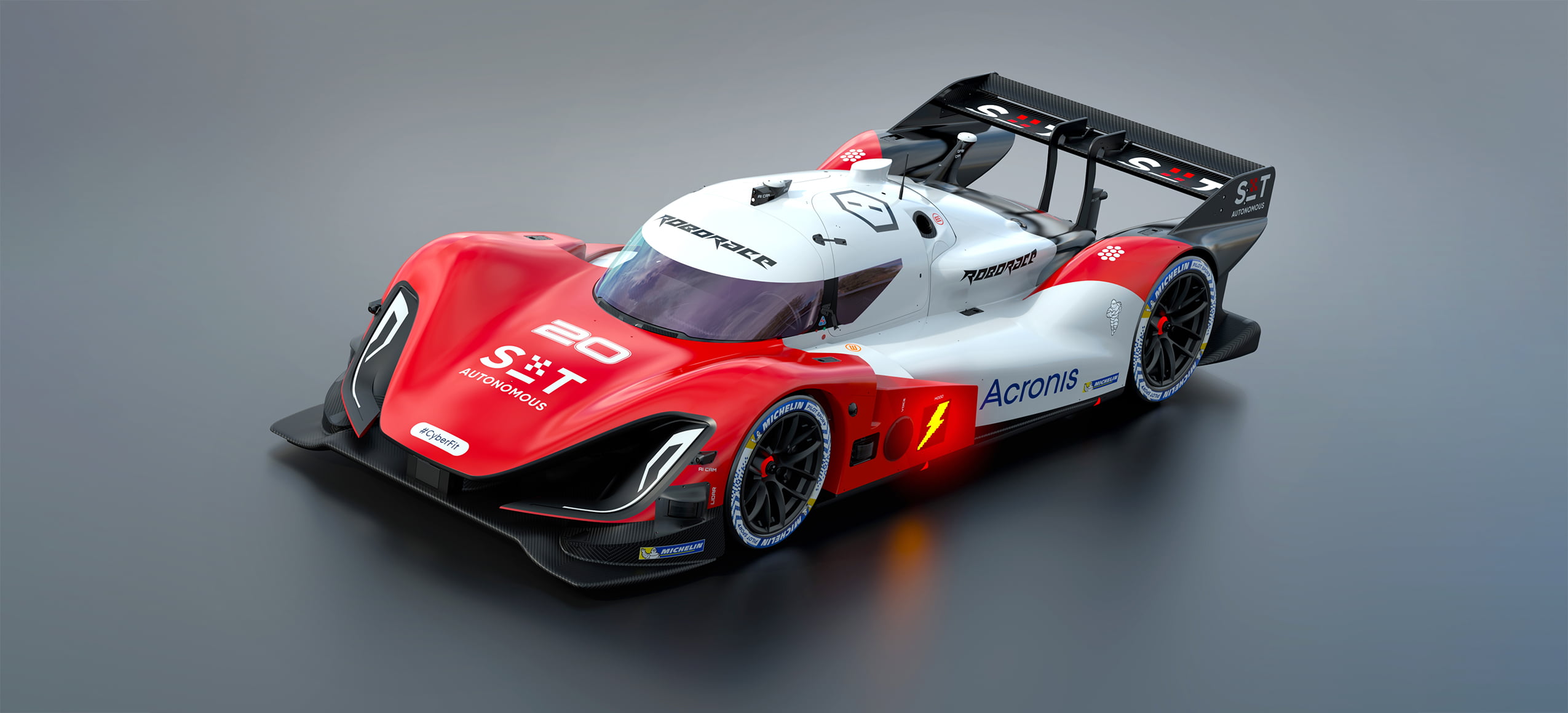}
\caption{Devbot 2.0 -- all-electric vehicle currently used in the Roborace competition (source: \url{https://roborace.com/})}
\label{roborace}
\end{figure}

Roborace \cite{roborace_site} is a global championship between autonomous cars. 
The hardware (the race cars) is the same for all participating teams; each gets access to an autonomous race car called Devbot 2.0, and develops software to drive it in races, in a completely autonomous way. Each season sees changes in the goals and rules, and the introduction of new conditions.

The race takes place on a circuit. 
The cars start at a designated spot on a starting grid and have to accomplish a given number of laps faster than the competing teams.
The competitors race independently and their racing time is compared after all teams have finished the race. 

No physical objects other than the competing race cars are present on a racetrack, but there can be virtual objects, of three kinds: static obstacles, loots and ghost cars. 
Cars get bonus time for collecting loots and penalty time for hitting obstacles or ghost cars. 
The total time is defined as the race time minus bonus time plus penalty time. 

\subsection{Roborace: use cases} \label{roborace_use_cases}

A specification of the Roborace system would include many use cases, such as:

\begin{itemize}
\item Race without obstacles
\item Avoid obstacles or stop
\item Update speed limit
\item Race with virtual obstacles
\item Race with virtual race cars
\item Move to pit
\item Perform an emergency stop
\item Perform a safe stop.
\end{itemize}

Let us pick the first of these, ``\textit{Race without obstacles}'', for further analysis. This subsection shows a typical use-case model for it; in the next subsection we will see how to integrate it into a more general OO framework.

A specification of the use case in the style introduced in section \ref{use_cases} is presented in table \ref{raceUC}.

\begin{small}
\begin{longtable}{
|p{0.17\textwidth}|p{0.76\textwidth}| }

\caption{A detailed description of the \textit{``race without obstacles''} use case}
\label{raceUC}
\\
 \hline
 Name & Race\_no\_obstacles \\
  \hline
 Scope & System  \\ 
  \hline
 Level & Business summary  \\ 
 \hline
 Primary actor & Race car Operator \\
 \hline
 Secondary actor & Roborace Operator \\
 \hline
 Context of use & Race car has to obey an instruction \\
 \hline
 Preconditions  & * Race car is on the racetrack grid\\
                & * Race car is not moving \\
                & * The global plan (trajectory and velocity profile) minimizing the race time is calculated \\
                & * The green flag is shown by the Roborace\\
 \hline
 Trigger & The system receives a request from the race car operator to start the race \\
 \hline
  Main success scenario & * The system calculates the local plan (path and velocity profile) during the race trying to follow the global plan as close as possible \\
                        & * The race car follows the local plan\\
                        & * After finishing the required number of laps the race car performs a safe stop  \\
\hline
Success guarantee & The race car has completed the required number of laps and stopped.\\
\hline
Extensions
 & A. The red flag received during the race \\
 & * The race car recalculates a global plan to perform an emergency stop \\
 & * The race car performs an emergency stop\\
 & B. The yellow flag is received during the race\\
 & * The system sets the speed limit according to the received value\\
 & * The race car finishes the race following the global trajectory and not exceeding the new speed limit\\
 & C. The difference between the calculated (desired) location and real (according to the sensors) location is more than a given threshold\\
 & * The race car recalculates a global plan to perform an emergency stop \\
 & * The race car performs an emergency stop\\
 \hline
Stakeholders 
& Race car Operator (requests the car to start the race) \\
and interests & Roborace Manager (sets the race goals and policies)\\
& Roborace Operator (shows the green, yellow, red flags)\\
 \hline

\end{longtable}
\end{small}

\subsection{Roborace: some requirements classes} \label{roborace_classes}

We now temporarily set use cases aside and consider what an object-oriented requirements model for Roborace would look like. Its classes would cover both:

\begin{itemize}

\item
Its environment, with such classes as \e{RACE_TRACK, MAP} and \e{OBSTACLE}.

\item
Components of the system, with such classes as \e{RACE_CAR, PLANNING_MODULE}, \e{CONTROL_MODULE} and \e{PERCEPTION_MODULE}.

\end{itemize}

Here is a sketch of two such classes, both belonging to the system.

\begin{lstlisting}[language=OOSC2Eiffel]
class RACE_CAR feature
    control_module: CONTROL_MODULE
    perception_module: PERCEPTION_MODULE
    planning_module: PLANNING_MODULE
    localization_and_mapping_module: LOCALIZATION_AND_MAPPING_MODULE
end

\end{lstlisting}

\begin{lstlisting}[language=OOSC2Eiffel]
class RACE_TRACK feature
    raceline: RACELINE
      -- Optimal raceline for the track
    map: MAP
      -- Coordinates of the bounding lines
end
\end{lstlisting}

\noindent During the development process, elements of the system’s functionality are assigned as features of respective modules’ classes and are enriched with contracts.

Below is an implementation of the requirement ``At every position on a raceline the speed in the velocity profile shall not exceed the maximum racecar's speed'': 


\begin{lstlisting}[language=OOSC2Eiffel]
class PLANNING_MODULE  feature
    car: RACE_CAR
    calculate_raceline (t: RACE_TRACK)
      --Calculate optimal raceline for a given racetrack
        do
        ensure
            across t.raceline.velocity_profile as rl all rl.item < car.max_speed end
        end
end
\end{lstlisting}

\subsection{Roborace: integrating the use cases into the object-oriented model} \label{OO_roborace_use_cases}

We now see how to express the use cases (section \ref{roborace_use_cases}) as part of the OO requirements (section \ref{roborace_classes}).

The ``race without obstacles'' use case, previously expressed in tabular format (Table \ref{raceUC}) becomes simply a routine \e{race_no_obstacles}  in the requirements class \e{ROBORACE_USE_CASES} sketched above. It relies on conditional expressions to consider the use case alternative flows. 

\begin{lstlisting}[language=OOSC2Eiffel]
race_no_obstacles 
    Note
        Callers: car_operator    
    require
        not car.is_moving 
        car.global_plan_is_calculated
        car.green_flag_is_up 
        car.is_on_starting_grid
    local local_plan: RACELINE
    do
        from --Sequence of system actions in use case main flow
        until car.race_is_finished or 
              car.red_flag_is_up or 
              car.location_error_is_detected 
        loop
            if car.yellow_flag_is_up then
              update_speed
            end
        local_plan := car.planning_module.calculate_local_plan
        car.control_module.move (local_plan.speed,
                                    local_plan.orientation)
        end
        if car.red_flag_is_up or car.location_error_is_detected then    
          emergency_stop
        else safe_stop end
    ensure
        not car.is_moving 
        car.is_in_normal_mode implies car.race_is_finished
    end
\end{lstlisting}

\noindent The \e{race_no_obstacles} routine relies on implementing the routines \e{update_speed}, \e{safe_stop}, and \e{emergency_stop}: the respective features are called inside the use case. 
These features are implementation of the respective use cases, and such dependency corresponds to \texttt{<<include>>} and \texttt{<<extend>>} relationships between use cases.

The \e{ROBORACE_USE_CASES} class is thus a collection of routines corresponding to the system's use cases. 

\begin{lstlisting}[language=OOSC2Eiffel]
class ROBORACE_USE_CASES feature
    car: RACE_CAR
	
    safe_stop 
        require
          ` car.is_in_normal_mode
        do
            car.control_module.safe_stop
        ensure
            not car_is_moving
        end

    emergency_stop
        require
            car.red_flag_is_up or car.location_error_is_detected
        do
            car.control_module.emergency_stop
        ensure
            not car.is_in_normal_mode
            not car.is_moving
        end


    update_speed
        require
            car.yellow_flag_is_up
        do
            car.update_max_speed (car.safe_speed)
        ensure
            car.current_max_speed = car.safe_speed
        end
			

    race_no_obstacles 
        do
          --implementation is listed above
        end
    
    avoid_obstacle_or_stop
        do
        end

    race_with_virtual_obstacles
        do
        end
    
    race_with_virtual_race_cars
        do
        end
    
    move_to_pit
        do
        end
    
end 
\end{lstlisting}

\subsection{Roborace: relation between use cases and test cases}
\label{use cases to test cases}

Use case stories define test cases for use cases \cite{jacobson2016usecase}.
 \e{ROBORACE_USE_CASE_STORIES} class inherits from \e{ROBORACE_USE_CASES} class. It includes a collection of routines corresponding to use case stories. 

When a use case takes the form of a routine with contracts, extracting use case stories from such a routine becomes a semi-automated task. 
For example, the \e{emergency_stop} use case accepts two options in its precondition -- (1) when the red flag is shown or (2) when a location error is detected.
These options map to the following use case stories written in the unified approach / with OO requirements:

\begin{lstlisting}[language=OOSC2Eiffel]
emergency_stop_red_flag_story
    require car.red_flag_is_up
    do emergency_stop end
	
emergency_stop_location_error_story
    require car.location_error_is_detected
    do emergency_stop end
\end{lstlisting}

\noindent These routines represent the two different paths through the \e{emergency_stop} use case, characterized by their preconditions.
The connection with the parent use case is visible because the stories call the routine encoding the use case.
The two routines must be exercised at least once with test input that meets their preconditions.

A similar analysis makes it possible to extract 5 use case stories from the \e{race_no_obstacles} use case:

\begin{itemize}
    \item three for each possible loop exit condition.
    \item one corresponding to the true antecedent of the implication in the second postcondition assertion.
    \item one corresponding to the true consequent and false antecedent of the said implication. 
\end{itemize}

\noindent The full collection of the extracted use case stories may be found in a publicly available repository \cite{roborace_spec}.

\subsection{Roborace: lessons from the example}

The ``main scenario'' of the \textit{``Race without obstacles''} use case (section \ref{roborace_use_cases}) provides a good illustration of the difference between contract-based and scenario-based specification (section \ref{logical}). As a specification, this scenario expresses, among other properties, that the system calculates a local plan and then follows it. 
It states this property in the form of a strict sequence of operations which, however, only covers some of the many possible scenarios.

It does list extensions, but only three of them, and does not reflect the many ways in which they can overlap. For example:
\begin{itemize}
    \item 

It can happen that the green flag is shown some time after the yellow flag; but the extensions do not even list the green flag.
\item
In the same way, the red flag can be shown after a yellow flag. 

\end{itemize}

\noindent An attempt to add extensions to cover all possibilities would have no end, as so many events may occur as to create a combinatorial explosion of possible sequencings.

One way out of this dead end would be to use temporal logic \cite{Pnueli}, which provides a finite way to describe a possibly infinite but constrained set of sequences of events or operations. The Design-by-Contract-based technique discussed in the present work relies on a different idea: use logical rather than sequential constraints. For each operation, we specify both:

\begin{itemize}
    \item The conditions it requires (precondition).
    \item The conditions it ensures (postcondition).
\end{itemize}

\noindent Sequential constraints become just a special case: we can express that \e{A} must come before \e{B} simply by defining a condition \e{C} as part of both the postcondition of \e{A} and the precondition of \e{B}. But the logic-based specification scheme covers many more possibilities than just this special case. 

In the example just mentioned, we state (in class \e{ROBORACE}) the constraint on raising the yellow and red flags:

\begin{lstlisting}[language=OOSC2Eiffel]
class ROBORACE feature
    raise_yellow_flag
        require
            green_flag.is_up
        do
        ensure
            yellow_flag.is_up
            not green_flag.is_up
            not red_flag.is_up
        end
    raise_red_flag
        require
            green_flag.is_up or yellow_flag.is_up
        do
        ensure
            red_flag.is_up
            not green_flag.is_up 
            not yellow_flag.is_up 
        end
end
\end{lstlisting}

\noindent Preconditions and postconditions apply to individual operations or events and cannot capture general environment constraints, such as the requirement that if the yellow flag is up cars should limit their speed to a dedicated ``safe speed''. For such requirements properties, we need contract elements of the third major kind, class invariants, as in the following extract from the specification of cars:

\begin{lstlisting}[language=OOSC2Eiffel]
class RACE_CAR feature
    green_flag_is_up: BOOLEAN
    yellow_flag_is_up: BOOLEAN
    red_flag_is_up: BOOLEAN
    safe_stop_activated: BOOLEAN
    max_speed: REAL	
    current_max_speed: REAL		
        -- Current speed limit
    safe_speed: REAL
        -- Safe speed limit
invariant
    yellow_flag_is_up implies current_max_speed = safe_speed
    green_flag_is_up implies current_max_speed = max_speed
    red_flag_is_up implies safe_stop_activated
end
\end{lstlisting}

\noindent This example is typical of how invariants capture fundamental consistency constraints. Almost every problem domain has such constraints, defining what is and is not possible. Any good requirements should include them. They have no place, however, in a  specification based solely on scenarios, which describe only examples of use, not the underlying invariant properties.

\section{Related Work}
\label{related}

\subsection{UML and SysML}
Use cases are an important modeling tool in UML \cite{UML}. \cite{Larman} illustrates use-case-driven requirements specification in UML and \cite{Overgaard} describes use-case patterns and blueprints for use case modeling in UML.

 UML makes it possible to treat use cases as objects, subject to specialization and decomposition. As noted in section \ref{scenario_classes}, the use-case classes described in this paper cover a different concept: a groups of related behaviors, pertaining to one or more data abstractions. UML use cases can have pre- and postconditions; in OO modeling as described in the present paper, pre- and postconditions (routine contracts) apply to individual operations. If a use case consists of a sequence of operations op$_{1}$, op$_{2}$, ... op$_{\textit{n}}$, with each op$_{\textit{i}}$ characterized by pre$_{\textit{i}}$ and post$_{\textit{i}}$, the pre- and postconditions of the use case are just pre$_{1}$ and post$_{\textit{n}}$. It is possible in UML to associate contracts with individual operations through  natural language or the OCL (Object Constraint Language) notation. 

SysML \cite{SysML}, an extended profile for UML, treats requirements as first class entities, establishing direct links between requirements and other software artifacts (such as tests). \cite{weilkiens2011systems} illustrates requirements specification process with SysML and \cite{Apvrille_SysML, xie2022sysml, waseem2018application} provide applications of SysML to all phases of software development. SysML does not provide semantics for requirements although it is possible to associate contracts with individual operations through natural language or the OCL notation. SysML and UML are standardized notations, rather than methodologies. 

\subsection{Use case modeling}

The Restricted Use Case Modeling approach \cite{yue2016practical, Yue} relies on a use case template and a set of restriction rules to reduce ambiguity of use case specification and facilitate transition to analysis models, such as UML class diagram and sequence diagram. The aToucan tool automates generation of UML class, sequence and activity diagrams \cite{yue2015atoucan}. The approach does not advocate extracting abstract properties from use cases and domain knowledge, such as time-ordering constraints and environment constraints.

Like the use case classes presented in section \ref{scenario_classes}, a Use Case Map (UCM) \cite{buhr1995use, amyot2005ucm} depicts several scenarios simultaneously.
UCMs represent use cases as causal sequences of responsibilities, possibly over a set of abstract components. 
In UCMs pre- and postconditions of use cases as well as conditions at selection points can be modeled with formal specification techniques such as ASM or LOTOS.  
UCMs specify properties of operations in relation with scenario sequences, rather than abstract properties of objects and operations.

Logical constraints, discussed in section \ref{logical} can also be formulated with state-based notations, such as Alloy \cite{jackson2012software}, Event-B \cite{abrial2010modeling, uc-b}, Abstract State Machines \cite{borger2010abstract} and Statecharts \cite{whittle2000generating, use-case_statecharts}. Some approaches \cite{uc-b, use-case_statecharts} apply formal modeling to specify use cases, but the formal specification is for an entire use case, not for its individual operations.

\subsection{Use cases as requirements}

There is disagreement among researchers about whether use cases are requirements. In the ICONIC process \cite{Rosenberg} use cases are requirements and constitute the main input for software design. Larman claims in \cite{Larman} that use cases are only part of requirements, constituting functional requirements, yet not all requirements. Other requirements artifacts include supplementary specification, glossary, vision and business rules. In our perspective, although use cases are an important source of requirements, they have to be thoroughly analyzed together with environment constraints before proceeding to software design.

\subsection{Contract-based approaches}  \label{contract_based}
To express properties of operations and their interactions, object-oriented requirements as discussed in this article rely on contracts rather than sequential ordering. Contracts also play a role in other requirements work, including approaches based on UML thanks to the Object Constraint Language, OCL. Larman's book on  UML and patterns \cite{Larman} specifies ``system operation contracts'' in natural language or OCL to express how system operations change the state of domain model objects. Soltana et al. \cite{soltana2014using, soltana2016model} apply OCL contracts to express operational legal requirements as policy models.

The present study does not include empirical evidence, in particular about the value of using contracts for requirements. Such evidence does appear in an empirical study by Briand et al. \cite{Briand2005, Briand2011AnEE}, which identified a positive effect from producing OCL contracts in requirements analysis. The effect is moderate, but the authors point out that participants lacked training in UML and OCL. 

Arnold et al. \cite{arnold2010modeling} formalize use cases as grammars of responsibilities. Abstract Constraint Language contracts (pre- and postconditions and invariants) capture constraints that scenarios' or responsibilities' execution poses on the system's state. In this approach a dedicated binding tool maps elements of a requirements model to elements of candidate implementation. The approach preserves the procedural nature of specification, since it organizes specification around scenarios and operations, rather than types of objects of the application's domain.

The SIRCOD approach \cite{FlorianPhD} provides a pipeline for converting natural language requirements to programming language contracts. The approach relies on the domain-specific language RSML for automating conversion from natural language to programming language. In the SOOR approach \cite{naumchev2019seamless, naumchev2016complete}, requirements are documented as software classes which makes them verifiable and reusable. Routines of those classes, called specification drivers, take objects to be specified as arguments and express the effect of operations on those objects with pre- and postconditions. The SIRCOD and SOOR approaches focus on translating existing requirements specifications to contracts expressed in a programming language, rather than extracting abstract requirements from scenarios.

\section{Discussion and conclusion} \label{discussion}

\subsection{Limitations}

The work presented here is a conceptual contribution to the area of requirements methodology, and has not  undergone any systematic empirical validation. The case study of section \ref{application} serves as a proof of concept on a significant ongoing project (but only one).

The absence of large-scale empirical studies makes it impossible to state any guarantees that the approach will increase productivity or decrease defects. The case for it is based instead on arguments
of a logical nature, in particular the observation that object-oriented technology with logical constraints is more general than scenario-based techniques and encompasses them as a special case. The next sections summarize these arguments.

The role of this article -- in the tradition of work arguing for specific approaches in software methodology -- is to provide the conceptual basis for studying OO versus scenario-based techniques. We do plan to conduct empirical studies to assess the actual consequences on actual projects. Studies conducted by others would be even more welcome. 

\subsection{Why OO requirements?}

 The idea of applying OO techniques to programming (their original focus since the appearance of the first object-oriented languages) is not controversial. Neither is their application to the next task up the abstraction level, software design. Moving up again one notch, to requirements, is not a new idea  \cite{coleman1994object, OOSC, Larman}; but OO requirements, unlike OO programming and OO design, are not widely practiced, for a large part because use cases and user stories have occupied the territory.
 
 As the discussion in this article argues, these techniques (if used as the principal form of requirements) are a step backward since scenarios, as a requirements technique, lack abstractness, generality and precision. One of their principal limitations is that they prescribe specific orderings of operations, which is generally too constraining, as there are simply too many such orderings to understand and state. Instead of specifying that operations must appear in specified orders, a more flexible approach specifies that:

\begin{itemize}
    \item 
    Prior to executing, each operation may require some properties.
    \item
    After executing, each operation will ensure some properties.
    \item
    Specifying these before-and-after properties is an insightful way of specifying the operations, and determines, as a byproduct, which scenarios are legal and which are not. 
\end{itemize}

\noindent A role remains for scenarios: to check that the more abstract logical specification does cover the specific cases (the most frequent usage patterns) that stakeholders have volunteered, as well as critical extreme cases that could lead to incorrect behavior. That role remains important, as long as we remember that:

\begin{itemize}
    \item The specification is not given by the scenarios.
    \item It is given instead by the logical constraints (pre- and postconditions, invariants) on operations (grouped, in an OO approach, into classes).
    \item The scenarios, while not providing the specification, provide \textit{examples} (often, important ones) of the specification, useful to validate that specification.
    \item It is part of the tasks of the requirements process to check that the scenarios respect the contracts (each operation works properly if its precondition is satisfied, ensures its postcondition, and preserves the invariant).
    
\end{itemize}

\noindent Applying OO principles to requirements helps producing more abstract and precise requirements due to the following mechanisms:

\begin{itemize}
    \item The approach motivates asking precise questions of the non-technical stakeholders to clarify delicate points. 
    \item Once a logical specification is produced, requirements engineer checks whether it corresponds to the stakeholders’ informal view, thus avoiding incorrect implementation due to misunderstandings.
    \item Since time-ordering constraints are more abstract than behavioral sequences, formulating them with contracts covers a wider range of possible scenarios.    
\end{itemize}

\subsection{Seamlessness}
Object-Oriented Analysis and Design (OOAD) approach relies on use cases along with object models as the techniques for requirements analysis \cite{pastor1997oo, rumbaugh1991object, ivarjacobson, Amyot2020, VonOlberg2022}. 
Unlike the Unified approach, in the OOAD approaches the process of software development is not seamless and involves several notations (natural language, UML, possibly formal languages). 
\cite{yue2015atoucan, Amyot2020, SoaviZMM21, VonOlberg2022} propose methods for conversion between notations.
However, the process of conversion is prone to errors and its outcome is never 100\% correct 
As a result, the source code cannot be statically verified against the requirements. 

As stakeholder requirements are formulated in natural language, a transition to programming language is inevitable at some point in the development process since software is written in a programming language. Seamless development makes the transitions as smooth as possible by encouraging the use of a single notation and a common set of concepts, embodied in the object-oriented paradigm. Transition to programming language as the notation during requirements analysis makes it possible to resolve ambiguities in requirements early in software development lifecycle.

\subsection{Future work}
One direction of future work is to apply the approach of unifying specifications, scenarios and tests to several case studies to provide more evidence for method validation. Another direction is to enrich the approach with tools that will improve its usability. Finally, we aim to perform a user study to evaluate the approach's usability and effectiveness.

\section{Conclusion}
\label{conclusion}
Two of the central problems of software engineering, as relevant to requirements as for design and implementation, are size and change. Software systems can be large and complex; they must adapt to modification. Any approach to software construction must be judged by its ability to help address these issues.

On the design and implementation side, object-oriented techniques meet this criterion: thanks to techniques of class-based modularity, information hiding, genericity and inheritance, they have shown that they can support the development of large, evolving systems.

This article has presented the application of object-oriented ideas to requirements, where we may expect that they yield the same benefits. The key result is that we do not need to treat object-oriented requirements as a competitor to other popular requirements techniques such as use cases, use case slices and user stories. Object-oriented decomposition is at a higher level of generality than such procedural techniques and encompasses them. More specifically:
\begin{itemize}
\item The core idea of object-oriented decomposition is to use classes as the basic modular unit.
\item The most intuitively appealing view of a class is that it describes a set of objects. 
For requirements, this concept of object is already rich in possibilities, since it makes it possible to describe, in a simple and natural way, concrete and abstract objects from a system's environment, such as a racetrack or a book catalogue (in the examples of this paper).
\item Beyond this initial view of objects, however, the full notion of object is more general: the most important characteristic of a class is that it gathers a group of operations (commands and queries) applicable to a given abstraction. Beyond environment objects, such an abstraction can describe a system concept, such as elements of a design pattern.
\item An object can also, in this spirit, describe a group of scenarios (use cases, user stories) relative to a set of abstractions, or a group of tests exercising these abstractions.
\end{itemize}

\noindent Applying these ideas results in a scheme that encompasses all the major requirements techniques in a general framework, with the advantages of conceptual consistency (as everything proceeds from a single overall idea, object technology) and unification of notations, and the potential of transferring the OO benefits of scalability and extendibility to the crucial discipline of requirements engineering.

\bibliographystyle{abbrv}
\bibliography{main}

\end{document}